\documentclass[12pt]{article}
\usepackage{subfiles}
\usepackage{amsmath,amssymb}
\usepackage{physics}
\usepackage{enumerate}
\usepackage{siunitx}
\usepackage{graphicx,caption}
\usepackage{slashed}
\usepackage{hyperref}
\usepackage{float}
\usepackage[top=25mm, bottom=30mm, left=30mm, right=30mm]{geometry}
\usepackage[style=phys,biblabel=brackets,chaptertitle=false,pageranges=false,eprint]{biblatex}

\numberwithin{equation}{section}

\begin{document}

\begin{flushright}
\hfill{KUNS-3037}
\end{flushright}
\begin{center}
\vspace{2ex}
{\Large \textbf{
Energy-Momentum Tensor and D-term of Baryons in Top-down Holographic QCD}}

\vspace*{5mm}
\textsc{Shigeki Sugimoto}$^{a,b,c}$\footnote{e-mail:
 \texttt{sugimoto@gauge.scphys.kyoto-u.ac.jp}}
 ~and~
\textsc{Taichi Tsukamoto}$^{a}$\footnote{e-mail:
 \texttt{taichi.t@gauge.scphys.kyoto-u.ac.jp}}

\vspace*{4mm}

\hspace{-0.5cm}
\textit{{$^a$
Department of Physics, Kyoto University, Kyoto 606-8502, Japan
}}\\
\textit{{$^b$
Center for Gravitational Physics and Quantum Information, Yukawa Institute for Theoretical
 Physics, Kyoto University, Kyoto 606-8502, Japan
}}\\
\textit{{$^c$
Kavli Institute for the Physics and Mathematics of the Universe (WPI),\\
 The University of Tokyo, Kashiwanoha, Kashiwa 277-8583, Japan
}}\\
\end{center}

\vspace*{.5cm}
\begin{abstract}

We study the energy-momentum tensor of a baryon in a top-down holographic QCD. In holographic QCD, the baryons are represented as solitons in a 5-dimensional gauge theory. We obtain the soliton solution by solving the equations of motion numerically. Using this result, the energy-momentum tensor and
related quantities such as the mass, mean square radii, and the D-term (druck term) are computed. The evaluated D-term is about $-2.05$, whose absolute value is significantly larger than that in the previous work \cite{Fujita:2022jus}.
\end{abstract}

\newpage
\tableofcontents

\section{Introduction}

The gravitational form factors are important quantities to probe the internal structures of hadrons.\footnote{See, e.g. \cite{Polyakov:2019lbq} for a review.} They are defined by the matrix elements of the energy-momentum tensor (EMT) and carry the information of the distributions of mass, spin, pressure, and shear force inside the hadron.
Recently, these quantities have been extracted from experimental data \cite{Kumano:2017lhr,Burkert:2018bqq,Kumericki:2019ddg,Burkert:2021ith,Kou:2021qdc,Wang:2022ndz} and triggered a lot of theoretical developments.
(See \cite{Polyakov:2019lbq,Burkert:2023wzr} and the references therein.)
In general, the evaluation of hadronic matrix elements in quantum chromodynamics (QCD), especially at low energies, is complicated because it requires serious analysis in a strongly coupled system involving the physics of bound states. Therefore, it would be worthwhile to investigate various methods to calculate these quantities and compare the predictions with each other and with experimental data.

The main goal of this paper is to report an improved numerical analysis of the gravitational form factors, in particular the so-called D-term for the nucleon, using a top down holographic QCD proposed in \cite{Sakai:2004cn,Sakai:2005yt}. The holographic description of QCD is obtained from a configuration of D-branes that realizes QCD in string theory. Since string theory contains gravity, it provides a natural framework to read off the gravitational form factors through the gravitational interactions of the hadrons.

The explicit procedure to obtain the gravitational form factors for baryons in this model was investigated in \cite{Fujita:2022jus} in detail and the value of the D-term was estimated. \footnote{See \cite{Fujii:2024rqd} for the recent study of the gravitational form factors for pion in this model.} It was pointed out that the gravitational form factors (in the leading order of the $1/N_c$ expansion) can be expressed as a sum over the contributions of glueball exchange diagrams, which was referred to as the glueball dominance. Furthermore, it was shown that in the zero momentum transfer limit, the gravitational form factors can be estimated by the EMT obtained in the meson effective theory. Based on these results, the value of the D-term was estimated to be around $-0.14$. However, the analysis in \cite{Fujita:2022jus} cannot be considered complete. One of the drawbacks was that the soliton configuration was obtained not by solving the equations of motion (EOMs), but by smoothly connecting the solutions near the center and the boundary.
In this paper, we improve this point by numerically solving the EOMs. It turns out that the value of the D-term is sensitive to the configuration of the gauge field in the intermediate region and it is crucial to accurately obtain the solution there. Our prediction of the value of the D-term is around $-2.05$. We also compute mean square radii, energy density, pressure, and shear force.

The organization of the paper is as follows. We start by reviewing the gravitational form factor in section \ref{GFF} and summarizing useful formulas. In section \ref{EMT}, we investigate the EMT obtained from the meson effective action in holographic QCD for the soliton solution under Witten's ansatz \cite{Witten:1976ck}.
Our numerical results are summarized in section \ref{sec:results}. The estimated values of the D-term and mean square radii are listed in Table \ref{tab:baryon-properties}. Section \ref{Conclusion} is devoted to the conclusion and discussion of possible future directions.

\section{Gravitational form factor}
\label{GFF}
\subsection{Definition of the gravitational form factors and the D-term}
In this subsection, we briefly review the definition of the gravitational form factors (GFFs) and the D-term based on \cite{Polyakov:2018zvc}.
\label{sec:GFF}
The GFFs are given by the matrix element of the EMT operator $\hat{T}_{\mu\nu}$. According to Lorentz invariance, it is written as 
\begin{equation}
    \begin{split}
         & \mel{p',s'}{\hat{T}_{\mu\nu}(x)}{p,s}\\
         & = \bar{u}\qty[A(t)\frac{P_{\mu}P_\nu}{m}+J(t)\frac{i(P_\mu\sigma_{\nu\rho}+P_\nu\sigma_{\mu\rho})\Delta^\rho}{2m}+D(t)\frac{\Delta_\mu\Delta_\nu-\eta_{\mu\nu}\Delta^2}{4m}+\bar{c}(t)\eta_{\mu\nu}]ue^{-i\Delta x}.
    \end{split}
    \label{AJD}
\end{equation}
for a spin-1/2 particle with mass $m$.\footnote{We use the Minkowski metric with the mostly-plus convention $\eta_{\mu\nu} = (-,+,+,+)$. } Here $\ket{p,s}$ is a one-particle state with momentum $p$ and helicity $s$, which satisfies the normalization condition $\ip{p',s'}{p,s} = 2p^0(2\pi)^3\delta(\vb*p'-\vb*p)\delta_{s',s}$. $P$, $\Delta$, and $t$ are defined as $P := (p+p')/2$, $\Delta := p'-p$, and $t := -\Delta^2$, respectively. $u(p,s)$ is the four-component spinor normalized as $\bar{u}(p,s)u(p,s) = 2m$. $A(t)$, $J(t)$, and $D(t)$ are the GFFs which are renormalization scale invariant scalar functions representing the inner structure of the particle. $\bar{c}(t)$ corresponds to the breaking of EMT conservation and it vanishes for the total EMT. It appears when the EMT is separated into some parts (in this paper, we will separate the EMT into the $SU(N_f)$ part and the $U(1)$ part). The number of form factors depends on the spin of the particle. For example, the $J$ term is absent for scalar particles. 

It is convenient to introduce the Breit frame $P = (E,\vb*0)$, $\Delta = (0,\vb*\Delta)$. In the Breit frame, the static EMT matrix elements are expressed as
\begin{align}
    \mel{p',s'}{\hat{T}^{00}(0)}{p,s} & = 2mE\qty[A(t)-\frac{t}{4m^2}\qty(A(t)-2J(t)+D(t))]\delta_{s's},\\
    \mel{p',s'}{\hat{T}^{0i}(0)}{p,s} & = 2mE\qty[J(t)\frac{(-i\vb*\Delta\times\vb*\sigma_{s's})_i}{2m}],
    \label{J}\\
    \mel{p',s'}{\hat{T}^{ij}(0)}{p,s} & = 2mE\qty[D(t)\frac{\vb*\Delta_i\vb*\Delta_j-\delta_{ij}\vb*\Delta^2}{4m^2}]\delta_{s's}.
\end{align}
Here, $\sigma_{s's}^i = \chi^\dagger_{s'}\sigma^i\chi_s$ with the Pauli spinors $\chi_{s'}$ and $\chi_s$ in the respective rest frames normalized as $\chi^\dagger_{s'}\chi_s = \delta_{s's}$.
Using the Fourier transformation
\begin{equation}
     T_{\mu\nu}(\vb*x) := \int\frac{\dd^3\vb*\Delta}{(2\pi)^32E}\mel{p',s}{\hat{T}_{\mu\nu}(0)}{p,s}e^{-i\vb*\Delta\cdot\vb*x},
\end{equation}
 $A(t)$, $J(t)$, and $D(t)$ satisfy
\begin{align}
    \label{eq:A-def}
    A(t)-\frac{t}{4m}\qty(A(t)-2J(t)+D(t)) & = \frac{1}{m}\int\dd^3\vb*xT^{00}(\vb*x)e^{i\vb*\Delta\cdot\vb*x},\\
    \label{eq:J-def}
    \qty(\delta_{ij}+\qty(t\delta_{ij}+\vb*\Delta_i\vb*\Delta_j){\dv{t}})J(t)s^j & = \int\dd^3\vb*x \epsilon_{ijk}x^jT^{0k}(\vb*x)e^{i\vb*\Delta\cdot\vb*x},\\
    \label{eq:D-def}
    \qty(1+\frac{4}{3}t{\dv{t}} +\frac{4}{15}t^2\dv[2]{t})D(t) & = -\frac{2}{5}m\int\dd^3\vb*x\qty(\vb*x^i\vb*x^j-\frac{1}{3}\delta^{ij}\vb*x^2)T_{ij}(\vb*x)e^{i\vb*\Delta\cdot\vb*x}.
\end{align}
Here, $s$ dependence of $T^{\mu\nu}(\vb*x)$ is implicit and $s^i = (\sigma_{ss})^i$ in \eqref{eq:J-def} is the unit vector along the spin direction.

In the forward limit $t\to0$, we have
\begin{align}
    A(0) & = \frac{1}{m}\int\dd^3\vb*xT^{00}(\vb*x) = 1,\\
    J(0)s_i & = \int\dd^3\vb*x\epsilon_{ijk}x^jT^{0k}(\vb*x) = \frac{1}{2}s_i,\\
    D(0) & = -\frac{2}{5}m\int\dd^3\vb*x\qty(\vb*x^i\vb*x^j-\frac{1}{3}\delta^{ij}\vb*x^2)T_{ij}(\vb*x).
\end{align}
The value of $A(0)$ is fixed to 1. This corresponds to the mass of the particle being $m$. In the same way, $J(0)$ is fixed to 1/2, which reflects the fact that the particle has spin 1/2. On the other hand, the value of $D = D(0)$ is not determined in general. This $D$ is called the D term and reflects the internal force distribution of the particle. The D-term is as fundamental as the mass and spin of a particle and is often referred to as ``the last unknown global property" \cite{Polyakov:2018zvc}. One of our goals is to estimate the value of the D-term of the baryon.

The value of the D-term for the nucleon has been evaluated using experimental data \cite{Burkert:2018bqq,Kumericki:2019ddg,Burkert:2021ith,Kou:2021qdc,Wang:2022ndz,Goharipour:2025lep}, bag model \cite{Neubelt:2019sou}, lattice QCD \cite{Shanahan:2018nnv,Shanahan:2018pib,Hackett:2023rif}, light-cone QCD sum rules \cite{Anikin:2019kwi,Azizi:2019ytx,Dehghan:2025ncw}, 
Skyrme model and its generalization \cite{Cebulla:2007ei,Jung:2013bya,Kim:2012ts,GarciaMartin-Caro:2023klo,GarciaMartin-Caro:2023toa},
chiral quark soliton model \cite{Goeke:2007fp,Wakamatsu:2007uc}, and bottom up holographic QCD \cite{Mamo:2019mka,Mamo:2021krl,Mamo:2022eui}. 
It is expected to be in the range $-5<D<-1$. 
For details, see \cite{Polyakov:2018zvc} and \cite{Cao:2024zlf}.
\subsection{Spherically symmetric case}
\label{sec:D-spherical}
In a spherically symmetric system, the stress tensor $T_{ij}$ can be decomposed into a trace part and a traceless part
\begin{equation}
\label{eq:EMT-symmetric}
    T_{ij}(\vb*x) = p(r)\delta_{ij}+s(r)\qty(n_in_j-\frac{1}{3}\delta_{ij}),
\end{equation}
where $p(r)$ is the pressure and $s(r)$ is the shear force inside the particle.
Substituting the spherically symmetric stress tensor \eqref{eq:EMT-symmetric} into \eqref{eq:D-def} and solving the differential equation, the form factor $D(t)$ can be expressed as 
\begin{equation}
\label{eq:D(t)-symmetric}
     D(t) = 16\pi m\int\dd rr^4s(r)\frac{j_2\qty(\sqrt{-t}r)}{tr^2},
\end{equation}
where $j_2(x)$ is the spherical Bessel function.
Taking the forward limit $t\to0$, we have
\begin{equation}
\label{eq:D-s}
    D = -\frac{16\pi}{15}m\int\dd rr^4s(r).
\end{equation}
The conservation law $\partial_\mu T^{\mu\nu}=0$ yields a constraint between $p$ and $s$,
\begin{equation}
\label{eq:conservation}
    p'(r)+\frac{2}{3}s'(r)+\frac{2}{r}s(r) = 0.
\end{equation}
From this equation, it can be shown that the pressure $p$ satisfies the von-Laue condition \cite{laue1911dynamik}
\begin{equation}
\label{eq:von-Laue}
    \int\dd rr^2p(r) = 0.
\end{equation}
This implies that the pressure has positive and negative regions.
Using \eqref{eq:conservation} and integration by parts, we can express the D-term in terms of $p(r)$ instead of $s(r)$,
\begin{equation}
\label{eq:D-p}
    D = 4\pi m\int\dd rr^4p(r).
\end{equation}
Note that \eqref{eq:D-s} and \eqref{eq:D-p} are equivalent only for the total EMT. For the separated EMT ($SU(N_f)$ part and $U(1)$ part, for example), the non-conserving term $\bar{c}$ also appears on the right-hand side of \eqref{eq:D-p}. 

\section{The energy-momentum tensor in holographic QCD}
\label{EMT}
\subsection{The model}
\label{sec:SS-model}
 The model we consider is the top-down holographic QCD model introduced in \cite{Sakai:2004cn,Sakai:2005yt}. It is a holographic dual of $SU(N_c)$ QCD with $N_f$ massless quarks realized in a system of $N_c$ D4-branes compactified on $S^1$ of radius $M_\mathrm{KK}^{-1}$ with $N_f$ pairs of D8-$\mathrm{\overline{D8}}$ branes. The holographic dual description is obtained by replacing the D4-brane with the corresponding supergravity background \cite{Witten:1998zw}. This gives a system of $N_f$ D8-branes embedded in this background, and closed and open strings correspond to glueballs and mesons, respectively.

 The low energy effective action of open strings on the D8 branes is the $U(N_f)$ Yang-Mills theory with the Chern-Simons term on a 1+4 dimensional curved background spacetime,
\begin{equation}
\label{eq:action-SSmodel}
    S = -\frac{\kappa}{2}\int\dd^4 x\dd z\tr\qty(h(z)\mathcal F_{\mu\nu}^2+2k(z)\mathcal F_{z\mu}^2)+\frac{N_c}{24\pi^2}\int\omega_5(\mathcal A).
\end{equation}
We take the Kaluza-Klein mass $M_\mathrm{KK} = 1$ for simplicity.
Here $\mu = 0,1,2,3$ are 1+3 dimensional Lorentz indices, and $z$ is the coordinate of the fifth space-like dimension. $\mathcal A = \mathcal A_\mu\dd x^\mu+\mathcal A_z\dd z$ is the $U(N_f)$ gauge field and $\mathcal F:=\dd \mathcal A+i\mathcal A\wedge\mathcal A$ is its field strength. We can decompose $\mathcal A$ into $SU(N_f)$ part and $U(1)$ part,
\begin{equation}
    \mathcal A = \mathcal A^{SU(N_f)}+\frac{1}{\sqrt{2N_f}}1_{N_f}\mathcal A^{U(1)}.
\end{equation}
 $\omega_5(\mathcal A)$ is the Chern-Simons 5-form of $\mathcal A$. The constant $\kappa$ is related to the 't Hooft coupling $\lambda$ (at the scale of $M_{\rm KK}$) as
\begin{equation}
    \kappa = \frac{N_c\lambda}{216\pi^3}.
\end{equation}
$h(z)$ and $k(z)$ are given by
\begin{equation}
    h(z) = (1+z^2)^{-\frac{1}{3}},\,k(z) = 1+z^2.
\end{equation}
The EOM for the action \eqref{eq:action-SSmodel} is 
\begin{equation}
\label{eq:EOM_Sakai-Sugimoto}
\begin{split}
    -2\kappa \qty(\mathcal{D}_\nu(h(z){\mathcal F}^{\nu\mu})+\mathcal D_z\qty(k(z)\mathcal F^{z\mu})) & =\frac{N_c}{16\pi^2}\epsilon^{\mu\nu\rho\sigma z}({\mathcal F}_{\rho\sigma}\mathcal F_{\nu z}+\mathcal F_{\nu z}\mathcal F_{\rho\sigma})\\
    -2\kappa \mathcal D_\mu(k(z)\mathcal F^{\mu z}) & =\frac{N_c}{32\pi^2}\epsilon^{\mu\nu\rho\sigma z}\mathcal F_{\mu\nu}\mathcal F_{\rho\sigma} 
\end{split}
\end{equation}
where $\mathcal D_\mu = \partial_\mu+i\comm{\mathcal A_\mu}{\cdot}$ and $\mathcal D_z = \partial_z+i\comm{\mathcal A_z}{\cdot}$ are the covariant derivatives, and $\epsilon_{\mu\nu\rho\sigma}$ is a 4-dimensional Levi-Civita symbol with $\epsilon_{0123z} = +1$.
In this model, the fluctuations of the $U(N_f)$ gauge field represent massless pions and massive (axial) vector mesons, while baryons are obtained as topological solitons in 1+4 dimensional spacetime \cite{Sakai:2004cn,Sakai:2005yt,Hata:2007mb}. The soliton corresponding to a baryon is an instanton on the 4-dimensional space which carries a non-trivial topological number given by the second Chern number
\begin{equation}
\label{eq:Chern-number}
    B = \frac{1}{8\pi^2}\int\dd^3\vb*x\dd z\epsilon^{0ijk z}\tr(\mathcal F_{ij}\mathcal F_{kz}) = -\frac{1}{8\pi^2}\int\dd^3\vb*x\dd z\epsilon_{0ijk z}\tr(\mathcal F_{ij}\mathcal F_{kz}).
\end{equation}
Here $\vb*x = (x^1,x^2,x^3)$ and $i,j,k = 1,2,3$. 
This is related to the baryon number in the Skyrme model. 
\subsection{Soliton solution}
We take $N_f = 2$ in this paper for simplicity.
In order to construct a soliton solution with baryon number $B=1$, we set Witten's ansatz \cite{Witten:1976ck}\footnote{See, e.g. \cite{Cherman:2011ve,Bolognesi:2013nja,Panico:2008it,Rozali:2013fna,Suganuma:2020jng,Hori:2023fxq} for closely related numerical analyses of solitons using Witten's ansatz.}
\begin{align}
    \label{eq:Witten-ansatz_Aw}
    \mathcal A_z^{SU(2)} & = A_z(r,z)n^a\tau_a\\
    \label{eq:Witten-ansatz_Ai}
    \mathcal A_i^{SU(2)} & = A_r(r,z)n_in^a\tau_a +\frac{\Phi_1(r,z)}{r}\qty(\tau_i-n_in^a\tau_a)-\frac{\Phi_2(r,z)+1}{r}\epsilon_{ija}n_j\tau_a\\
    \label{eq:Witten-ansatz_A0}
    \mathcal A_0^{U(1)} & = a_0(r,z),\\
    \label{eq:Witten-ansatz_else}
    \mathcal A_0^{SU(2)} & = \mathcal A_z^{U(1)} = \mathcal A_i^{U(1)} = 0,
\end{align}
where $r = |\vb*x|$ is the radial coordinate and $\vb*n := \vb*x/r$ is the unit vector of the 3-dimensional space. $\tau_a:=\sigma_a/2,\,(a = 1,2,3)$ are the generators of $SU(2)$ Lie algebra.
This ansatz is invariant under $SO(3)$ ``spatial rotation'' combined with the global $SU(2)$ transformation
\begin{equation}
\label{eq:rotation}
\begin{split}
    \mathcal A_{z}^{SU(2)}(\vb*x,z) & = U(R)\mathcal A_{z}^{SU(2)}(R^{-1}\vb*x,z)U(R)^\dagger,\\
    \mathcal A_{i}^{SU(2)}(\vb*x,z) & = U(R){R_i}^j\mathcal A_j^{SU(2)}(R^{-1}\vb*x,z)U(R)^\dagger,
\end{split}\quad R\in SO(3)
\end{equation}
where $U(R)$ is the $SU(2)$ representation of $R$.
Notice that this ansatz has a residual $U(1)$ gauge transformation
\begin{equation}
    \mathcal A_M(x)^{SU(2)}\to e^{i\lambda(r,z)\vb*n\cdot\vb*\tau}\qty(\mathcal A_M^{SU(2)}(x)-i\partial_M)e^{-i\lambda(r,z)\vb*n\cdot\vb*\tau}\quad ( M = 0,1,2,3,z).
\end{equation}
Under this transformation, $\Phi:=\Phi_1+i\Phi_2$, $A_r$, and $A_z$ transform as
\begin{equation}
\begin{split}
\Phi\to e^{-i\lambda}\Phi,\,\,
    A_\alpha\to A_\alpha-\partial_\alpha\lambda\quad(\alpha = r,z).
\end{split}
\end{equation}
Therefore, $(A_r,A_z)$ and $\Phi$ can be considered as a $U(1)$ gauge field and a $U(1)$ charged scalar, respectively, in the $r$ - $z$ plane. It is convenient to introduce the $U(1)$ field strength and covariant derivative on the $r$ - $z$ plane
\begin{equation}
    F_{\alpha\beta}:=\partial_\alpha A_\beta-\partial_\alpha A_\beta,\,D_\alpha\Phi:=(\partial_\alpha-iA_\alpha)\Phi.
\end{equation}
and the 1+4 dimensional Yang-Mills theory is reduced to 2 dimensional Abelian-Higgs theory. The Chern number \eqref{eq:Chern-number} is written as the winding number of the vortex
\begin{equation}
\label{eq:Chern-number-2d}
    B = -\frac{1}{2\pi}\int\dd r\dd z\qty(F_{rz}(1-\abs{\Phi}^2)+2\Im(D_r\Phi^\dagger D_z\Phi))=:\int\dd r\dd z\mathrm{ch}_2(r,z),
\end{equation}
which only depends on the boundary values of $\Phi$ and $A_\alpha$ as
\begin{equation}
\label{eq:Baryon-number_2dim}
    B = \int\dd r\qty[A_r+\Im(\Phi^\dagger D_r\Phi)]_{z = -\infty}^{z = \infty}-\int_{-\infty}^{\infty}\dd z\qty[A_z+\Im(\Phi^\dagger D_z\Phi)]_{r\to0}^{r\to\infty}.
\end{equation}
With this ansatz, the action \eqref{eq:action-SSmodel} is written in terms of $S = -\int\dd tM_\mathrm{sol}$, where $M_\mathrm{sol}$ is the classical mass of the soliton
\begin{equation}
\label{eq:Classic-Action}
\begin{split}
    M_{\mathrm{sol}} & = 4\pi\kappa\int\dd r\dd z\qty(\frac{r^2}{2}kF_{rz}^2+h|D_r\Phi|^2+k|D_z\Phi|^2+\frac{h}{2r^2}(1-\abs{\Phi}^2)^2)\\
    & -4\pi\kappa\int\dd r\dd z\frac{r^2}{2}\qty(h(\partial_ra_0)^2+k(\partial_za_0)^2)\\
    & +4\pi\kappa\gamma\int\dd r\dd z a_0(F_{rz}(1-\abs{\Phi}^2)+2\Im(D_r\Phi^\dagger D_z\Phi)),
\end{split}
\end{equation}
where $\gamma = \frac{N_c}{16\pi^2\kappa}$. The Euler-Lagrange equations of \eqref{eq:Classic-Action} are
\begin{align}
\label{eq:EOM-Witten_Phi}
    (hD_r^2+D_z(kD_z))\Phi+\frac{h}{r^2}(1-\abs{\Phi}^2)\Phi & = i\gamma\qty(\partial_ra_0D_z\Phi-\partial_za_0D_r\Phi),\\
    \partial_z(r^2kF_{zr})+2h\Im(\Phi^\dagger D_r\Phi) & = \gamma\partial_za_0(1-\abs{\Phi}^2),\\
    \partial_r(r^2F_{rz})+2k\Im(\Phi^\dagger D_z\Phi) & =-\gamma\partial_ra_0(1-\abs{\Phi}^2),\\
\label{eq:EOM-Witten_a_0}
    (h\partial_r(r^2\partial_r)+r^2\partial_z(k\partial_z))a_0 & = -\gamma\qty(F_{rz}(1-\abs{\Phi}^2)+2\Im(D_r\Phi^\dagger D_z\Phi)).
\end{align}
These equations can also be derived from substituting the ansatz into the original EOM \eqref{eq:EOM_Sakai-Sugimoto}.

\subsection{The energy-momentum tensor in Witten's ansatz}
In AdS/CFT correspondence, the proper method to calculate the expectation value of the
EMT has been established \cite{Balasubramanian:1999re,deHaro:2000vlm}. The application of this method to a baryon in holographic QCD was investigated in \cite{Fujita:2022jus}, and it was shown that the value of the D-term can be calculated from the classical 1+3 dimensional EMT defined as
\begin{equation}
\begin{split}
    T^{\mu\nu} & :=\eval{\frac{2}{\sqrt{-g}}\fdv{S}{g_{\mu\nu}}}_{g = \eta} = 2\kappa\int\dd z\tr(k\mathcal F^{z\mu}\mathcal F^{z\nu}+h\mathcal F^{\mu\rho}{\mathcal F^\nu}_\rho-\frac{1}{4}\eta^{\mu\nu}\qty(h\mathcal F_{\rho\sigma}^2+2k\mathcal F_{z\rho}^2)).
\end{split}
\label{emt2}
\end{equation}
Substituting the ansatz \eqref{eq:Witten-ansatz_Aw} to \eqref{eq:Witten-ansatz_else}, we have the spherically symmetric EMT 
\begin{align}
    T^{00}(\vb*x) & = \epsilon(r),\\
    T^{0i}(\vb*x) & = 0,\\
    T^{ij}(\vb*x) & = p(r)\delta_{ij}+s(r)\qty(n_in_j-\frac{1}{3}\delta_{ij}),
\end{align}
where the energy density $\epsilon(r)$, the pressure $p(r)$, and the shear force $s(r)$ are
\begin{align}
\begin{split}
    \epsilon(r) & = \epsilon^{SU(2)}(r)+\epsilon^{U(1)}(r) = \int\dd z\epsilon_z^{SU(2)}(r,z)+\int\dd z\epsilon_z^{U(1)}(r,z),\\
    \epsilon_z^{SU(2)}(r,z) & = \kappa\qty(\frac{1}{2}kF_{rz}^2+\frac{1}{r^2}(h|D_r\Phi|^2+k|D_z\Phi|^2)+\frac{h}{2r^4}(1-\abs{\Phi}^2)^2),\\
    \epsilon_z^{U(1)}(r,z) & = \kappa\frac{1}{2}(h(\partial_ra_0)^2)+k(\partial_za_0)^2),
    \end{split}\\
    \begin{split}
        p(r) & = p^{SU(2)}(r)+p^{U(1)}(r),\\
        p^{SU(2)}(r) & = \frac{\kappa}{3}\int\dd z\qty(-\frac{1}{2}kF_{rz}^2+\frac{1}{r^2}(h|D_r\Phi|^2-k|D_z\Phi|^2)+\frac{h}{2r^4}(1-\abs{\Phi}^2)^2),\\
        p^{U(1)}(r) & = \frac{\kappa}{3}\int\dd z\qty(\frac{1}{2}h(\partial_ra_0)^2+\frac{3}{2}k(\partial_za_0)^2),
    \end{split}\\
    \begin{split}
        s(r) & =  s^{SU(2)}(r)+s^{U(1)}(r),\\
        s^{SU(2)}(r) & = \kappa\int\dd z\qty(kF_{rz}^2+\frac{1}{r^2}(h|D_r\Phi|^2-k|D_z\Phi|^2)-\frac{h}{r^4}(1-\abs{\Phi}^2)^2),\\
            s^{U(1)}(r) & = - \kappa\int\dd zh(\partial_ra_0)^2.
    \end{split}
    \end{align}

The classical mass of the soliton is calculated as
\begin{equation}
\label{eq:mass}
    M_\mathrm{EMT}=M^{SU(2)}+M^{U(1)} = 4\pi\int\dd rr^2\qty(\epsilon^{SU(2)}(r)+\epsilon^{U(1)}(r)).
\end{equation}
It is easy to show that $M_\mathrm{sol} = M_\mathrm{EMT}$ using the $U(1)$ part Gauss-law equation \eqref{eq:EOM-Witten_a_0}.

As introduced in section \ref{sec:GFF}, we can derive the D-term in two ways, from the pressure $p$ \eqref{eq:D-p} and the shear force $s$ \eqref{eq:D-s}
\begin{align}
    D_p & = D_p^{SU(2)}+D_p^{U(1)} = 4\pi M_\mathrm{sol}\int\dd rr^4(p^{SU(2)}(r)+p^{U(1)}(r))\\
    D_s & = D_s^{SU(2)}+D_s^{U(1)} = -\frac{16\pi}{15} M_\mathrm{sol}\int\dd rr^4(s^{SU(2)}(r)+s^{U(1)}(r)).
\end{align}
$D_p$ and $D_s$ must be equal due to the conservation law \eqref{eq:conservation}. 
\subsection{The baryon mean square radii}
\label{sec:radii}
There are several ways to estimate the size of the soliton. One way is the mean square radius of the energy density
\begin{equation}
    \expval{r^2}_\epsilon:=\frac{1}{\int\dd^3\vb*x\epsilon(r)}\int\dd^3\vb*xr^2\epsilon(r) = \frac{1}{M_\mathrm{sol}}\int\dd^3\vb*xr^2\epsilon(r).
\end{equation}
In the spherically symmetric case, it has been argued that the radial force $F_r$ 
\begin{equation}
    F_{r}:=T_{ij}n^in^j = p(r)+\frac{2}{3}s(r)
\end{equation}
is always positive in a stable system \cite{Perevalova:2016dln} (see also \cite{Polyakov:2018zvc}). The mean square radius of $F_r$ 
\begin{equation}
    \expval{r^2}_{\mathrm{mech}}:=\frac{\int\dd^3\vb*xr^2 F_r}{\int\dd^3\vb*x F_r}.
\end{equation}
is called the mechanical radius.

\section{Analysis and Results}
\label{sec:results}
In this section, we show our results for a soliton solution with $B=1$ obtained by solving the EOMs
\eqref{eq:EOM-Witten_Phi}-\eqref{eq:EOM-Witten_a_0} numerically.  
The parameters $M_\mathrm{KK}$ and $\lambda$ are taken to be $949\,\si{MeV}$ and $16.65$, respectively, which are the values used in \cite{Sakai:2004cn,Sakai:2005yt} to reproduce the pion decay constant $f_\pi$ and the mass of the $\rho$ meson. We use the coordinate $w:= \arctan{z}$ instead of $z$ for the calculation. In this coordinate, we have $k(w)=1$, $h(w) = (\cos{w})^{-\frac{4}{3}}$, and the boundary corresponding to $z\to\pm\infty$ is $w = \pm w_\mathrm{max} = \pm\frac{\pi}{2}$. The cutoff of $r$ is set to $R_\mathrm{max} = 5\pi$ in the $M_\mathrm{KK} = 1$ unit, and this is physically $3.3\,\si{fm}$. The $r$ - $w$ plane is first reduced to $w\geq0$ by adopting the $\mathbb{Z}_2$ symmetry associated with $w\to-w$, and discretized to $30\times 300$ square lattice with spacing $\Delta = \pi/60$ or $0.01\,\si{fm}$. EOMs \eqref{eq:EOM-Witten_Phi}-\eqref{eq:EOM-Witten_a_0} are solved by the Gauss-Seidel method. For more details, see the Appendix \ref{appendix:BC}.

Figure \ref{fig:epsilon-wr} are the plots of $4\pi r^2\epsilon_w(r,w):= 4\pi r^2{\dv{z}{w}}(\epsilon_z^{SU(2)}+\epsilon_z^{U(1)})$ and $\mathrm{ch}_2(r,w)$ \eqref{eq:Chern-number-2d} on the $r$ - $w$ plane. The peak of $\epsilon_w$ is located at $w\neq0$. This is because $\epsilon_w$ includes the factor $h(w)$, which is an increasing function for $w>0$ and amplifies the $w \neq 0$ region of the energy density. In contrast, the center of $\mathrm{ch}_2$, which does not depend on $h(w)$, stays on the $w = 0$ axis.

\begin{figure}[H]
\centering
    \begin{minipage}{0.45\columnwidth}
        \centering
        \includegraphics[width = 7cm]{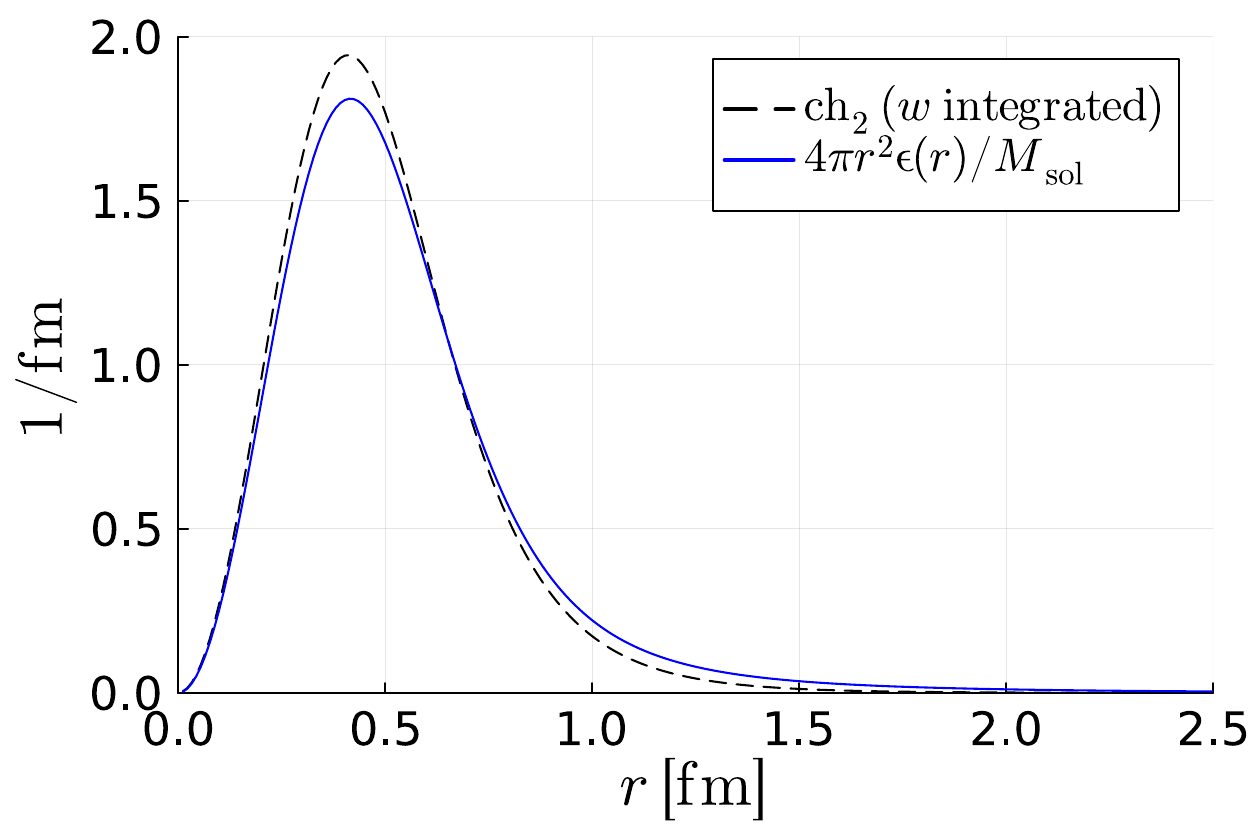}
    \end{minipage}
    \begin{minipage}{0.45\columnwidth}
        \centering
        \includegraphics[width = 7cm]{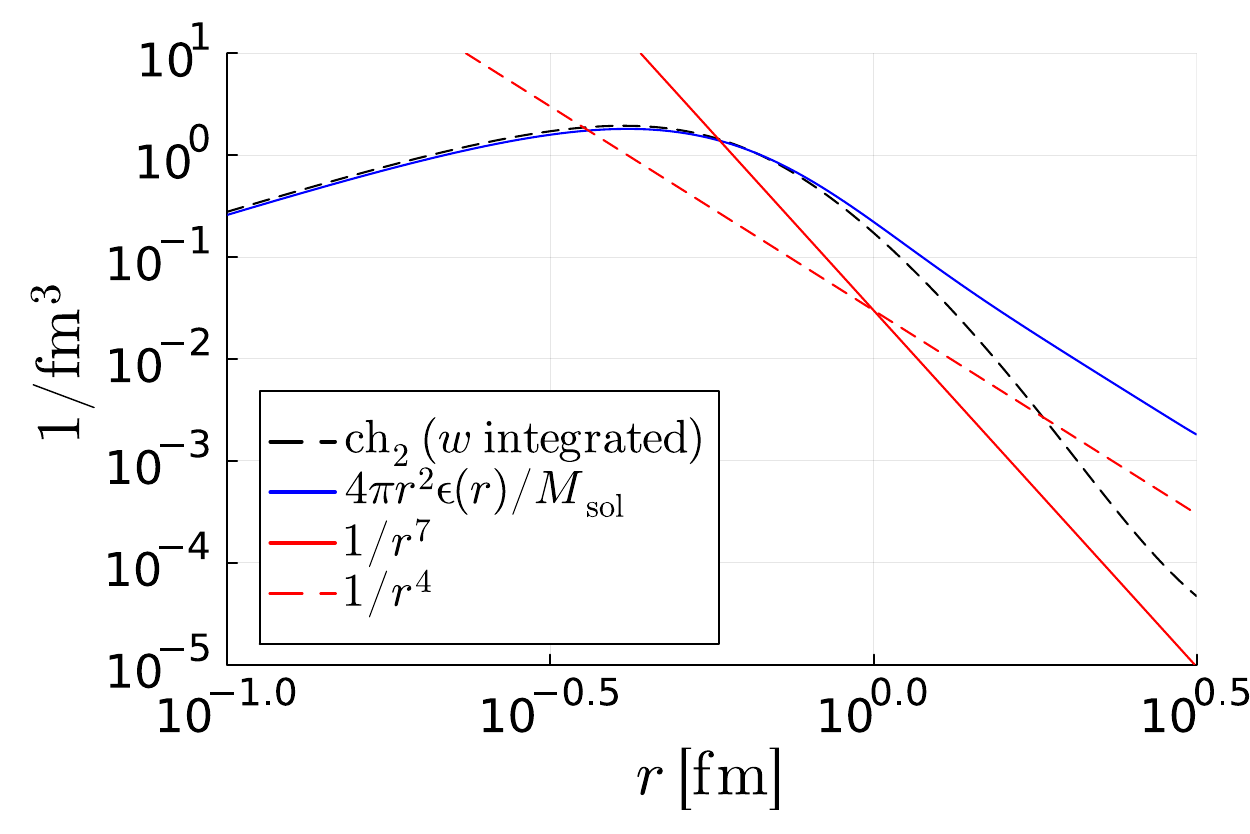}
    \end{minipage}
    \caption{Energy density and Chern number density along the $r$ direction. The black dashed line is $\mathrm{ch}_2(r,w)$ integrated along the $w$ axis. The blue solid line is $4\pi r^2\epsilon(r)$ normalized with $M_\mathrm{sol}$. The right figure is plotted in the log-scale. The red solid and dashed lines are $1/r^7$ and $1/r^4$, respectively.}
    \label{fig:epsilon}
\end{figure}

\begin{figure}[ht]
\centering
\begin{minipage}{0.45\columnwidth}
    \centering
    \includegraphics[width = 7cm]{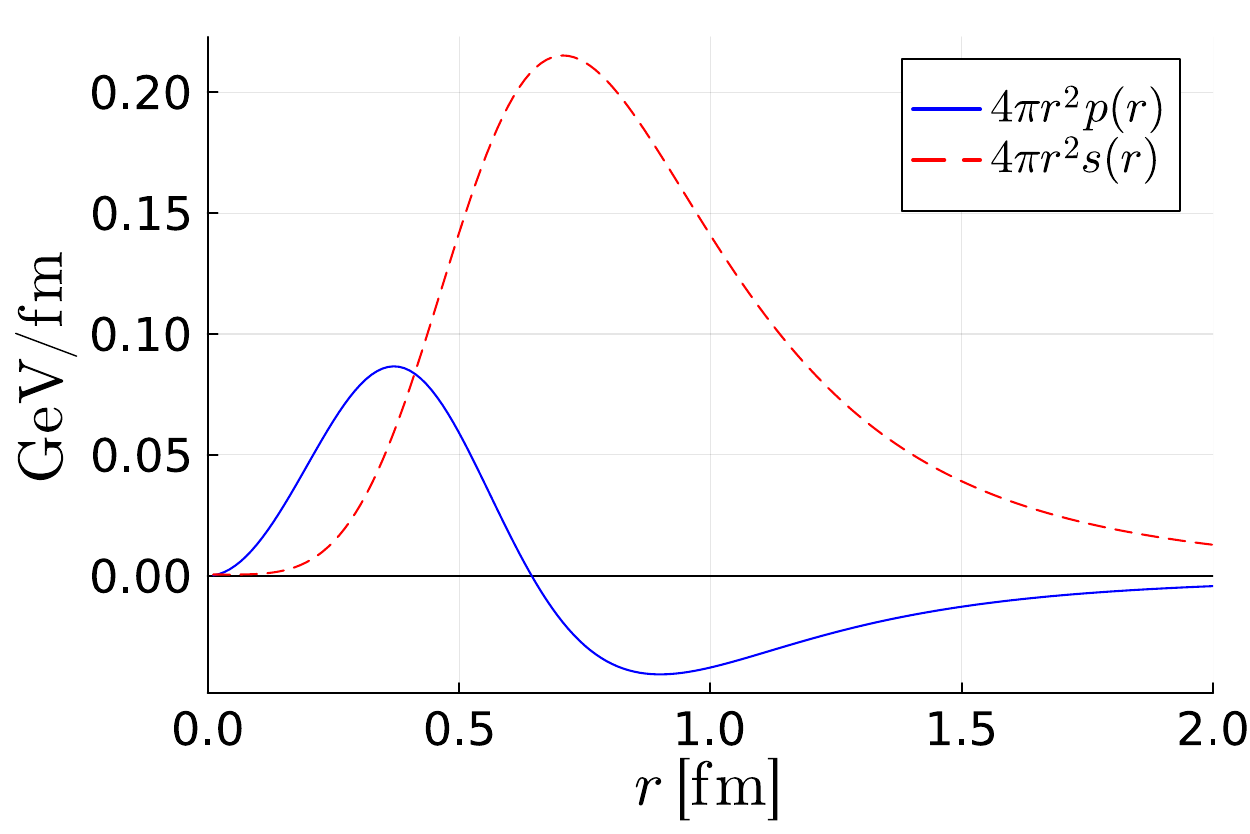}
    \caption{$p(r)$ and $s(r)$ multiplied with $4\pi r^2$. The blue solid line is $p(r)$ and the red dashed line is $s(r)$.}
    \label{fig:p-s}
\end{minipage}
~~~
\begin{minipage}{0.45\columnwidth}
    \centering
    \includegraphics[width = 7cm]{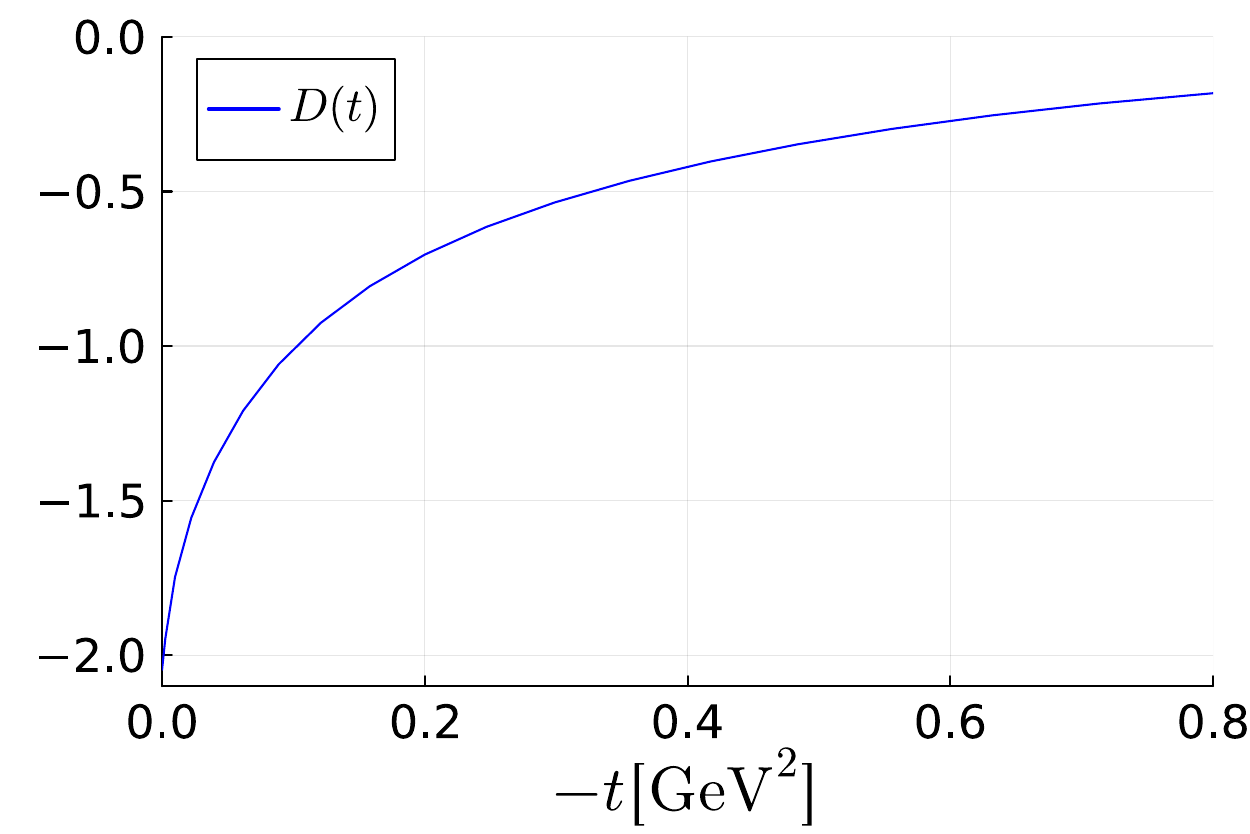}
    \caption{$t$ dependence of $D(t)$. The graph is derived from \eqref{eq:D(t)-symmetric} for $t\neq0$, and $D(0)$ is fixed to $D_s = -2.05$.}
    \label{fig:D(t)}
\end{minipage}
\end{figure}

Integrating Figure \ref{fig:epsilon-wr} along the $w$ axis, we obtain Figure \ref{fig:epsilon}. In the outer region $r\gtrsim 1.0\,\si{fm}$, $r^2\epsilon(r)$ behaves as $r^{-4}$, while $\mathrm{ch}_2(r):=\int\dd w\mathrm{ch}_2(r,w)$ decreases as $r^{-7}$. This is consistent with the asymptotic behavior in large $r$ derived analytically in \cite{Cherman:2011ve} and the EMT of the Skyrme model \cite{Cebulla:2007ei}.
The pressure $p(r)$ and the shear force $s(r)$ defined in \eqref{eq:EMT-symmetric} are shown in Figure \ref{fig:p-s}. 
$p(r)$ is positive in $r\leq0.64\,\si{fm}$ and negative outside. This behavior is consistent with experimental results \cite{Burkert:2018bqq,Wang:2022ndz,Goharipour:2025lep} and calculations in other models \cite{Neubelt:2019sou,Shanahan:2018nnv,Anikin:2019kwi,Dehghan:2025ncw,Cebulla:2007ei,Jung:2013bya,GarciaMartin-Caro:2023toa,Goeke:2007fp,Kim:2012ts,Mamo:2019mka,Mamo:2021krl,Mamo:2022eui}. To check the conservation law \eqref{eq:conservation}, we evaluate
\begin{equation}
    \int_0^\infty\dd r\abs{p'+2s'+\frac{2}{r}s}\simeq 1.4\,\si{MeV/fm^3}.
\end{equation}
This is about $1\,\%$ of $p(0)\simeq 114\,\si{MeV/fm^3}$, and it is small enough within the numerical precision.

Table \ref{tab:baryon-properties} is the list of global properties and the sizes of the baryon evaluated in our numerical calculation along with the value of D-term in \cite{Fujita:2022jus}. The classical mass of the baryon \eqref{eq:mass} is $M_{SU(2)} = 950\,\si{MeV}$, $M_{U(1)} =232\,\si{MeV}$, and $M_\mathrm{sol} = 1.18\,\si{GeV}$ in total. This value agrees well with the value obtained in \cite{Hori:2023fxq}, although this is $416\,\si{MeV}$ smaller than the value evaluated in \cite{Hata:2007mb}. The values of D-term are $D_p = -2.06$ and $D_s=-2.05$. The difference between $D_p$ and $D_s$ is less than $1\,\%$, and agrees with good precision. As mentioned in section \ref{sec:D-spherical}, the individual values of the $SU(2)$ part and $U(1)$ part of $D_p$ and $D_s$ do not agree because $D_p^{SU(2)}$ and $D_p^{U(1)}$ involve the contribution of a non-conserving term $\bar{c}$. The $t$-dependence of $D(t)$ can be obtained from \eqref{eq:D(t)-symmetric}, and the result is shown in Figure \ref{fig:D(t)}. $D(t)$ is negative anywhere, and converges to $0$ at large $t$. The slope at $t = 0$ is infinitely steep, since $\dv{t}D(t)$ is proportional to $\int\dd rr^6s(r)$ at small $t$, which diverges in the chiral limit.

The two mean radii defined in section \ref{sec:radii} are $\expval{r^2}_\epsilon = (0.662\,\si{fm})^2$ and $\expval{r^2}_\mathrm{mech} = (0.938\,\si{fm})^2$. These values are comparable to other recent results \footnote{See e.g. \cite{Goharipour:2025yxm} for an overview.}.

To compare our numerical results with the previous calculation in \cite{Fujita:2022jus}, we use $D_s$ because the D-term was calculated from the traceless part of the stress tensor in \cite{Fujita:2022jus}. We find qualitatively similar behavior such that the $SU(2)$ part is negative and $U(1)$ is positive, being negative in total in both cases. However, the absolute value of the $SU(2)$ part from our numerical result is about 4 times larger than that obtained in \cite{Fujita:2022jus}, making the total D-term more negative.
Note that the contribution from the $SU(2)$ part of $D_s$ is much enhanced compared to that of \cite{Fujita:2022jus}, while the $U(1)$ part is not that changed.
One possible reason for this is that, as pointed out in \cite{Fujita:2022jus}, the main contribution of the $SU(2)$ part is from the deviation of $SU(2)$ gauge fields from the self-dual instanton solution for the flat spacetime.
Since the soliton configuration used in \cite{Fujita:2022jus}
approaches the self-dual instanton in the small $w$ region, the effect of the breaking of the self-dual condition in this region was not properly included.
For the $U(1)$ part, on the other hand, the main contribution to the D-term is captured by the approximation in \cite{Fujita:2022jus} relatively well.

\begin{table}[ht]
    \centering
    \begin{tabular}{c|cccc|cc}
         & $M_\mathrm{sol}$ & $D_p$ & $D_s$ & $D$ in \cite{Fujita:2022jus} & $\expval{r^2}_\epsilon$ &  $\expval{r^2}_\mathrm{mech}$\\
         \hline
         results & $1.18\,[\si{GeV}]$ & $-2.06$ & $-2.05$ & $-0.140$ &$(0.662)^2\,[\si{fm}^2]$ &  $(0.938)^2\,[\si{fm}^2]$\\
         $SU(2)$ part & $950\,[\si{MeV}]$ & $-3.42$ & $-2.54$ & $-0.685$ & $(0.692)^2\,[\si{fm}^2]$ & $(1.38)^2\,[\si{fm}^2]$\\
         $U(1)$ part & $232\,[\si{MeV}]$ & $1.36$ & $0.489$ & $0.543$ &$(0.529)^2\,[\si{fm}^2]$ &  $(0.267)^2\,[\si{fm}^2]$\\
         \end{tabular}
    \caption{Mass, D-term and mean square radii.}
    \label{tab:baryon-properties}
\end{table}
\begin{figure}[ht]
\centering
    \begin{minipage}{0.45\columnwidth}
        \centering
        \includegraphics[width =7cm]{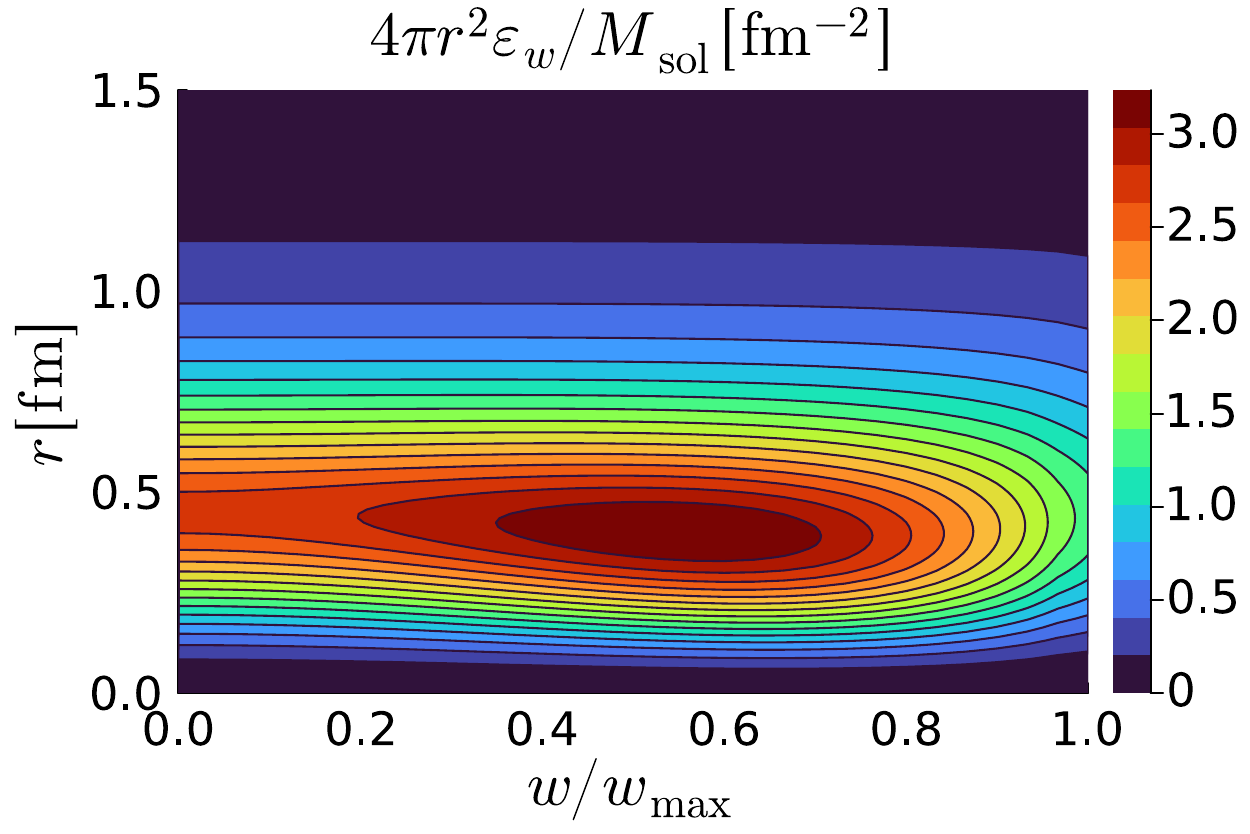}
    \end{minipage}
    \begin{minipage}{0.45\columnwidth}
        \centering
        \includegraphics[width = 7cm]{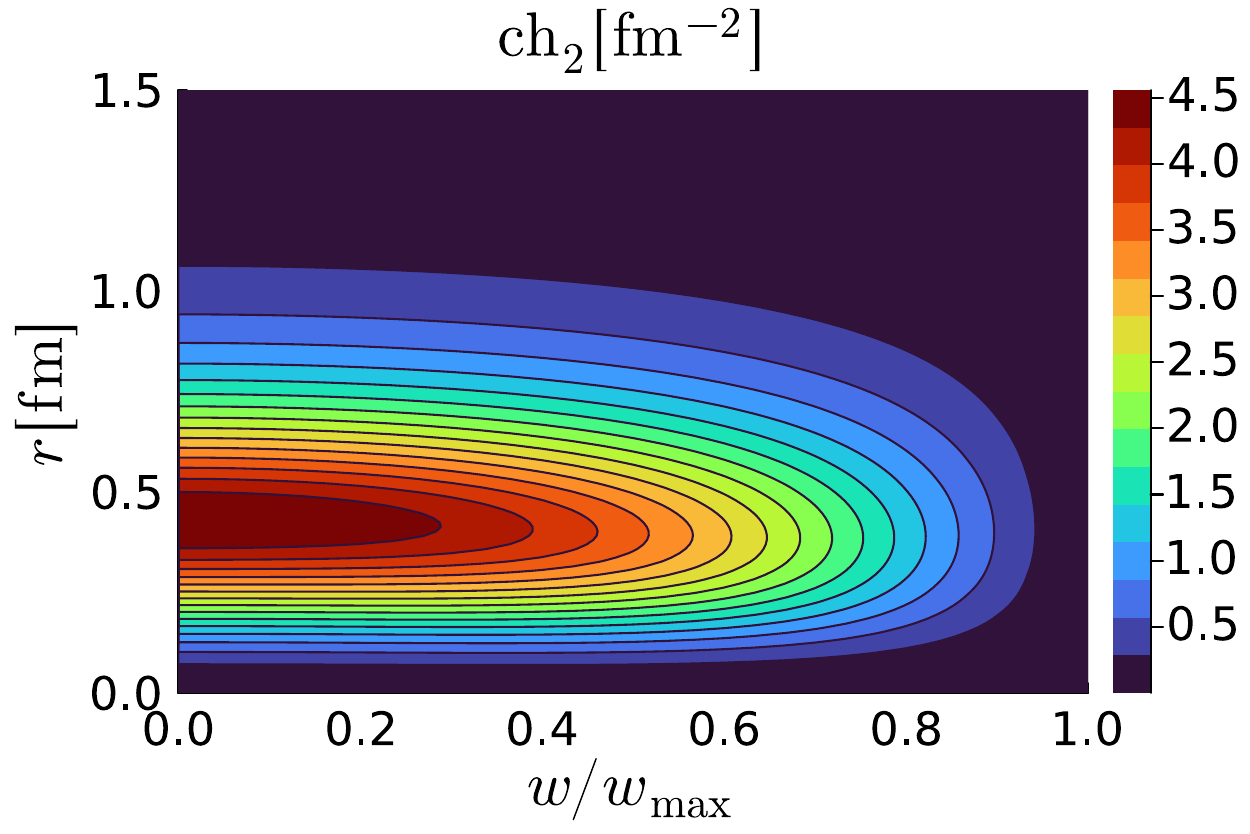}
    \end{minipage}
    \caption{Energy density and Chern number density of the baryon on $r$ - $w$ plane. The plot range is restricted to $0\leq w\leq w_\mathrm{max}=\pi/2$, which corresponds to $0\leq z<+\infty$.}
    \label{fig:epsilon-wr}
\end{figure}

\section{Conclusions}
\label{Conclusion}

In this paper, we investigated the gravitational form factors of a baryon using a top-down holographic description of QCD proposed in \cite{Sakai:2004cn,Sakai:2005yt,Hata:2007mb}.
In this framework, a baryon is described as a topologically stable solitonic gauge configuration in a 5-dimensional gauge theory. In order to obtain a classical solution corresponding to a baryon, we used Witten's ansatz and solved the equations of motion numerically. Using this numerical solution, the energy-momentum tensor was evaluated and the quantities such as energy density, pressure, and shear force were extracted. The results are summarized in Table \ref{tab:baryon-properties} and Figures \ref{fig:epsilon-wr}--\ref{fig:D(t)}. In particular, the D-term turned out to be significantly larger in absolute value compared to the previous result in \cite{Fujita:2022jus}. 

Our analysis is not complete and there are several directions to improve the calculation. First of all, our analysis is based on a static classical solution of the soliton corresponding to the baryon. It would be interesting to combine our numerical analysis with the quantization of the fluctuations around the soliton solution investigated in \cite{Hata:2007mb,Panico:2008it,Hashimoto:2008zw,Cherman:2011ve}. An important ingredient is the spin of the baryon, which can be introduced by quantizing the collective coordinates related to the rotational degrees of freedom. Once it is included, it will be possible to analyze the $J(t)$ form factor for nucleon (see (\ref{AJD}) and (\ref{J})) and the gravitational form factors for higher spin baryons (e.g. $\Delta$ resonance). (See \cite{Kim:2020lrs} for a work in this direction.\footnote{See also \cite{Pefkou:2021fni,Alharazin:2022wjj,Fu:2022rkn,Dehghan:2023ytx} for studies of GFFs for the $\Delta$ resonance using different approaches.}) The analysis of the electromagnetic form factors done in \cite{Hashimoto:2008zw} can also be improved by using our numerical analysis. Secondly, the model used in this paper describes QCD with massless quarks and the effect of the current quark mass is neglected. There are some proposals to add the quark masses in this system \cite{Aharony:2008an,Hashimoto:2008zw}. Since the effect of the quark mass drastically changes the IR behavior of some quantities, such as $\frac{d}{dt}D(t)|_{t\rightarrow 0}$ mentioned in section 4, it would be worthwhile to include it in the analysis of the form factors. For the study of hyperons, it will be crucial to include the strange quark mass to obtain realistic predictions. (See \cite{Hashimoto:2009st})
Thirdly, our analysis is based on the expression of the EMT in (\ref{emt2}). As it was shown in \cite{Fujita:2022jus}, (\ref{emt2}) is obtained as an approximation that is valid when the momentum transfer is smaller than the mass scale of the lightest glueball, which is of order 1 GeV. Therefore, for $t\sim
{\cal O}(1)\,\mbox{GeV}^2$ or higher, the glueball propagators (or, in other words, the bulk to boundary propagator) should be taken into account properly.
In addition, the analysis in the holographic description is done in the supergravity approximation, in which the $\alpha'$ corrections (corresponding to the $1/\lambda$ corrections) as well as quantum gravity effects (corresponding to the $1/N_c$ corrections) are neglected. It would be nice if these corrections could be incorporated.

\section*{Acknowledgement}
We are especially grateful
to Y. Hatta for valuable comments on a draft of this paper.
The work of SS was supported by the
JSPS KAKENHI (Grant-in-Aid for Scientific Research (B)) grant number JP24K00628
and MEXT KAKENHI (Grant-in-Aid for Transformative Research Areas A “Extreme Universe”) grant number 21H05187.
The work of TT was supported by JST BOOST, Grant Number JPMJBS2407.

\appendix
\section{Boundary conditions}
\label{appendix:BC}
In this appendix, we summarize the boundary conditions we used to solve the EOM \eqref{eq:EOM-Witten_Phi} - \eqref{eq:EOM-Witten_a_0}.
Under the ansatz \eqref{eq:Witten-ansatz_Aw} - \eqref{eq:Witten-ansatz_else}, the 5-dimensional gauge field $\mathcal A$ is reduced to real and complex scalars $a_0$ and $\Phi$, and a $U(1)$ gauge field $(A_z,A_r)$ on a 2-dimensional $z$-$r$ plane. Replacing $z$ with $w=\arctan(z)$, the boundaries of this plane are $r = 0$, $w = \pm\frac{\pi}{2}$, and $r \to \infty$.

From the regularity of $\mathcal A$, the boundary condition at $r=0$ are
\begin{equation}
    \Phi = -i,\,A_w = 0,\,\partial_rA_r = 0,\,\partial_ra_0 = 0,\quad(r=0).
\end{equation}
Because of the divergent factor $h(w)=\cos(w)^{-\frac{4}{3}}$, $\mathcal F_{\mu\nu}$ must vanish at $w = \pm\frac{\pi}{2}$ to make energy finite. Therefore, $\mathcal A_\mu$ is pure gauge at $w=\pm\frac{\pi}{2}$. We choose a gauge such that $\mathcal A_\mu \to 0$ at this boundary in this paper. Therefore we have
\begin{equation}
    \Phi = -i,\,A_r = 0,\,a_0 = 0,\quad\qty(w=\pm\frac{\pi}{2}).
\end{equation}
Similarly, $\mathcal F_{\mu\nu}$ and $\mathcal F_{w\mu}$ should be $0$ at $r\to\infty$. However, we cannot let $\mathcal A\to0$ as the boundary condition at $r\to\infty$. This is because we have set $\mathcal A_\mu \to 0$ at $w = \pm\frac{\pi}{2}$, and the baryon number \eqref{eq:Baryon-number_2dim} is now
\begin{equation}
    B = \eval{-\frac{1}{2\pi}\int_{-\frac{\pi}{2}}^{\frac{\pi}{2}}A_w\dd w}_{r\to\infty} = \eval{-\frac{1}{2\pi}\int_{-\frac{\pi}{2}}^\frac{\pi}{2}\Im(\Phi^\dagger\partial_w\Phi)\dd w}_{r\to\infty}.
\end{equation}
This leads to the non-trivial behavior of $A_w$ and $\Phi$ at $r\to\infty$. We take a simple asymptotic behavior
\begin{equation}
\label{eq:infinity}
    \Phi \to ie^{-2iw},\,A_w \to -2,\,A_r\to0,\,a_0\to0,\quad(r\to\infty).
\end{equation}

In the chiral limit, it is known that the EMT of a baryon decays with a power law as $r\to\infty$. Thus, merely using \eqref{eq:infinity} with the finite cutoff is not appropriate. Because of this, we set the asymptotic behavior of the fields following the discussion in \cite{Cherman:2011ve,Bolognesi:2013nja} as 
\begin{equation}
    A_w+2 = \order{1/r^2},\,|\Phi|^2 -1 = \order{1/r^4},
\end{equation}
at $r=R_\mathrm{cutoff}$. For other fields, we take the value at $r\to\infty$ as the boundary condition on $r= R_\mathrm{cutoff}$, since they converge faster than $1/r^5$.

\printbibliography[title = References]
\end{document}